\documentclass[reprint,superscriptaddress,amsmath,amssymb,aps,prl]{revtex4-2}

\usepackage{graphicx}
\usepackage{dcolumn}
\usepackage{bm}
\usepackage{mathrsfs}
\usepackage{amsmath}
\usepackage{hyperref}
\usepackage{sidecap}
\usepackage{textcomp}
\usepackage{amssymb}
\usepackage{multirow,booktabs}
\usepackage{tabularx}

\begin{document}

\title{Revealing spin-flip two-level systems using ultra-thin film superconducting resonators}

	\author{Zi-Qing Huang}
	\thanks{These authors contributed equally to this work.}
	\affiliation{CAS Key Laboratory of Quantum Information, University of Science and Technology of China, Hefei, Anhui 230026, China}
	\affiliation{CAS Center for Excellence in Quantum Information and Quantum Physics, University of Science and Technology of China, Hefei, Anhui 230026, China}
	
	\author{Shu-Kun Ye}
	\thanks{These authors contributed equally to this work.}
	\affiliation{CAS Key Laboratory of Quantum Information, University of Science and Technology of China, Hefei, Anhui 230026, China}
	\affiliation{CAS Center for Excellence in Quantum Information and Quantum Physics, University of Science and Technology of China, Hefei, Anhui 230026, China}
	
	\author{Yong-Qiang Xu}
	\affiliation{CAS Key Laboratory of Quantum Information, University of Science and Technology of China, Hefei, Anhui 230026, China}
	\affiliation{CAS Center for Excellence in Quantum Information and Quantum Physics, University of Science and Technology of China, Hefei, Anhui 230026, China}
	
	\author{Tian-Yi Jiang}
	\affiliation{CAS Key Laboratory of Quantum Information, University of Science and Technology of China, Hefei, Anhui 230026, China}
	\affiliation{CAS Center for Excellence in Quantum Information and Quantum Physics, University of Science and Technology of China, Hefei, Anhui 230026, China}
	
	\author{Tian-Yue Hao}
	\affiliation{CAS Key Laboratory of Quantum Information, University of Science and Technology of China, Hefei, Anhui 230026, China}
	\affiliation{CAS Center for Excellence in Quantum Information and Quantum Physics, University of Science and Technology of China, Hefei, Anhui 230026, China}
	
	\author{Bao-Chuan Wang}
	\affiliation{CAS Key Laboratory of Quantum Information, University of Science and Technology of China, Hefei, Anhui 230026, China}
	\affiliation{CAS Center for Excellence in Quantum Information and Quantum Physics, University of Science and Technology of China, Hefei, Anhui 230026, China}
	
	\author{Xiang-Xiang Song}
	\email{songxx90@ustc.edu.cn}
	\affiliation{CAS Key Laboratory of Quantum Information, University of Science and Technology of China, Hefei, Anhui 230026, China}
	\affiliation{CAS Center for Excellence in Quantum Information and Quantum Physics, University of Science and Technology of China, Hefei, Anhui 230026, China}
	\affiliation{Suzhou Institute for Advanced Research, University of Science and Technology of China, Suzhou, Jiangsu 215123, China}
	
	\author{Hai-Ou Li}
	\affiliation{CAS Key Laboratory of Quantum Information, University of Science and Technology of China, Hefei, Anhui 230026, China}
	\affiliation{CAS Center for Excellence in Quantum Information and Quantum Physics, University of Science and Technology of China, Hefei, Anhui 230026, China}
	\affiliation{Hefei National Laboratory, University of Science and Technology of China, Hefei, Anhui 230088, China}
	
	\author{Guang-Can Guo}
	\affiliation{CAS Key Laboratory of Quantum Information, University of Science and Technology of China, Hefei, Anhui 230026, China}
	\affiliation{CAS Center for Excellence in Quantum Information and Quantum Physics, University of Science and Technology of China, Hefei, Anhui 230026, China}
	\affiliation{Hefei National Laboratory, University of Science and Technology of China, Hefei, Anhui 230088, China}
	
	\author{Gang Cao}
	\email{gcao@ustc.edu.cn}
	\affiliation{CAS Key Laboratory of Quantum Information, University of Science and Technology of China, Hefei, Anhui 230026, China}
	\affiliation{CAS Center for Excellence in Quantum Information and Quantum Physics, University of Science and Technology of China, Hefei, Anhui 230026, China}
	\affiliation{Hefei National Laboratory, University of Science and Technology of China, Hefei, Anhui 230088, China}

	\author{Guo-Ping Guo}
	\affiliation{CAS Key Laboratory of Quantum Information, University of Science and Technology of China, Hefei, Anhui 230026, China}
	\affiliation{CAS Center for Excellence in Quantum Information and Quantum Physics, University of Science and Technology of China, Hefei, Anhui 230026, China}
	\affiliation{Hefei National Laboratory, University of Science and Technology of China, Hefei, Anhui 230088, China}
	\affiliation{Origin Quantum Computing Company Limited, Hefei, Anhui 230088, China}
	\date{\today}
	\begin{abstract}
		Material disorders are one of the major sources of noise and loss in solid-state quantum devices, whose behaviors are often modeled as two-level systems (TLSs) formed by charge tunneling between neighboring sites. However, the role of their spins in tunneling and its impact on device performance remain highly unexplored. In this work, employing ultra-thin TiN superconducting resonators, we reveal anomalous TLS behaviors by demonstrating an unexpected increase in resonant frequency at low magnetic fields. Furthermore, a spin-flip TLS model is proposed, in which an effective spin-orbit coupling is generated by inhomogeneous local magnetic fields from defect spins. This mechanism mixes charge tunnelings and spin flips, quantitatively reproducing the observed frequency-field relationship and its temperature dependence. This work deepens the understanding of spin-dependent TLS behaviors, offering the possibility of magnetically engineering noise and loss in solid-state quantum devices.
	\end{abstract}
\maketitle
As the functional complexity of solid-state quantum devices increases, the requirements for reducing decoherence and noise become more stringent~\cite{blais2004Cavity, devoret2013Superconducting}. Material defects have been widely recognized as the primary source of these issues~\cite{muller2019understanding}. Therefore, understanding the mechanisms of defects is crucial for enabling large-scale quantum applications. To date, various models have been proposed to understand how material disorders influence device performance~\cite{paladino2014noise}. For example, it is generally accepted that decoherence and charge noise in superconducting resonators and charge qubits originate from the electric dipole bath formed by two-level systems (TLSs)~\cite{muller2019understanding,paladino2014noise,degraaf2020Twolevel}. Meanwhile, the flux noise in superconducting quantum interference devices (SQUIDs) or flux qubits originates from the magnetic dipole bath formed by spins in magnetic defects~\cite{kumar2016Origin, sendelbach2008Magnetism}. Recently, several studies have indicated that the charge noise~\cite{degraaf2018Suppression} and the quality factor~\cite{jayaraman2024Lossa} of quantum devices, initially thought to be dominated by TLSs~\cite{gao2008semiempirical, gao2008Experimental}, change after removing interface magnetic defects through surface treatment. This finding hints at a subtle connection between TLSs and spins in magnetic defects. However, the physical mechanism of the interplay between them remains highly unexplored.

To understand the spin dynamics, the behaviors of devices under different magnetic fields need to be systematically investigated~\cite{laforest2015Fluxvector, rower2023Evolution}. Unfortunately, the performance of devices commonly used to study TLSs, such as superconducting resonators or SQUIDs, degrades rapidly when external magnetic fields are increased to only several tens of milliteslas, because of quasiparticles (qp) generation in superconductors~\cite{healey2008Magnetic, abrikosov1961Contribution, tinkham1996Introduction,foshat2023Characterizing}. This prevents clear identification of spin-dependent responses, which brings challenges to directly revealing TLSs related to spins. Recently, superconducting resonators made from extremely thin films with a high critical magnetic field have been shown to maintain performance in in-plane magnetic fields up to several teslas~\cite{samkharadze2016HighKineticInductance, yu2021Magnetic, zollitsch2019Tuning}, providing a promising platform to overcome the challenges effectively.

In this work, we use 7.9 nm thick TiN thin film superconducting resonators to investigate their resonant frequencies as a function of external magnetic fields. Deviating from the quadratic decrease predicted by the BCS theory, the frequencies first increase and then decrease with increasing in-plane field. Moreover, as the temperature ($T$) increases, this anomalous frequency shift gradually disappears, consistent with the thermal saturation of TLSs. To quantitatively understand these observations, we propose a spin-flip TLS model in which tunneling electrons interact with spins in magnetic defects. This work provides valuable insights into the connection between magnetic defects and TLSs, which deepen the understanding of the spin-dependent behaviors of TLSs.

\begin{figure}[tp]
	\centering
	\includegraphics[width = 8.6 cm]{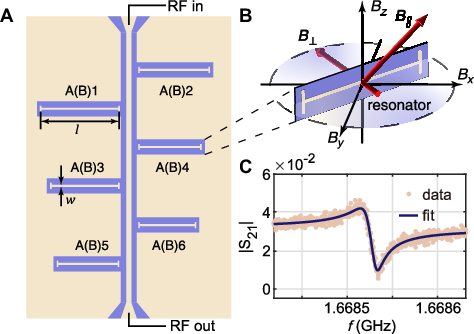}
	\caption[asd]{(A) Schematic diagram of the sample used in the work (not to scale), where the blue part represents the exposed substrate after etching the TiN thin film (yellow). (B) Schematic illustration of external magnetic fields with respect to the resonator, $B_{x,y,z}$ represent the axes of the magnet. (C) Measured $|S_{21}|$ from A1.}
	\label{fig1}
\end{figure}

The experiments are performed using a sample consisting of six nanowire resonators evenly coupled to a shared coplanar waveguide feedline (Fig.~\ref{fig1}A). We study two distinct samples, referred to as A and B, with resonator widths ($w$) of 300 and 500 nm, respectively. For each sample, six resonators are labeled as A(B)1--A(B)6, with various center conductor lengths ($l$). Magnetic fields are applied using a vector magnet (Fig.~\ref{fig1}B), and the resonant frequency ($f_r$) of each resonator is measured at a photon number $\langle n\rangle$ of $\sim5\times10^3$ using a vector network analyzer (Fig.~\ref{fig1}C). Such $\langle n\rangle$ ensures a high signal-to-noise ratio without causing nonlinear effects. The specific device parameters, fabrication procedure, and measurement setup can be found in Supplementary Material (SM).~\nocite{thomas2020Nonlinear,pompeo2008Reliable,pappas2011Two,stan2004Critical,nulens2023Catastrophic,degraaf2014Galvanically,shalibo2010Lifetime,sandberg2012Etch,rusevich2021electronica,kristen2024Giant}

\begin{figure}[tbp]
	\centering
	\includegraphics[width=8.6 cm]{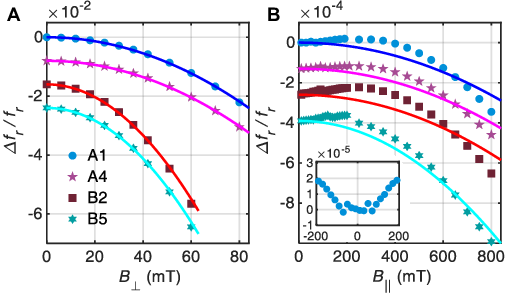}
	\caption[asd]{Measured $\Delta f_{r}/f_{r}$ as a function of (A) $B_{\perp}$ and (B) $B_{\parallel}$ under $T=10$ mK. Data sets have been offset for clarity. Inset in (B) shows $\Delta f_{r}/f_{r}$ of A1 when $B_{\parallel}$ is swept from 190 mT to $-190$ mT.}
	\label{fig2}
\end{figure}

Figure 2 shows the measured $f_r$ of four typical resonators as a function of magnetic field $B$ (results from the other eight resonators can be found in SM). Under a perpendicular magnetic field $B_{\perp}$, $f_r$ decreases rapidly. This is because the magnetic field breaks the Cooper pairs to generate qp, and increases the kinetic inductance of the resonator~\cite{abrikosov1961Contribution, zollitsch2019Tuning}. We use the fractional frequency shift $\Delta f_{r}/f_{r}$ to quantify the frequency shift. $\Delta f_{r}/f_{r}$ exceeds at least 2\% at $B_{\perp}=80$ mT, showing a quadratic behavior ($\Delta f_{r}^{\rm qp}/f_{r}=k B^2$) predicted by the standard BCS theory (solid curves in Fig.~\ref{fig2}A)~\cite{healey2008Magnetic, yu2021Magnetic}.

However, when a magnetic field is applied parallel to the resonator plane ($B_{\parallel}$), the effect of qp generation is much weaker. This is because the cross-sectional area of the resonator in the magnetic field is dominated by the thickness (7.9 nm) here, instead of the center conductor width (hundreds of nanometers) under $B_\perp$~\cite{samkharadze2016HighKineticInductance}. More importantly, the relatively high internal quality factor ($Q_i\sim10^5$) allows us to accurately detect small shifts in $f_r$. As shown in Fig.~\ref{fig2}B, when $B_{\parallel}$ increases to 800 mT, $\Delta f_{r}/f_{r}$ is less than 0.06\%. This enables us to distinguish signatures of other possible mechanisms beyond the qp generation, which may otherwise be masked by large $\Delta f_{r}^{\rm qp}/f_{r}$. Interestingly, we indeed find that the frequency evolution deviates from the quadratic fitting (solid curves in Fig.~\ref{fig2}B). More strikingly, $\Delta f_{r}/f_{r}$ shows an increasing trend for $B_\parallel<200$ mT, which is completely opposite to the predictions of the BCS theory.

This anomaly is reproducible since it appears in all 12 fabricated resonators. It is not caused by residual magnetic fields, as confirmed by scanning the field in the opposite direction (inset of Fig.~\ref{fig2}B). We further rule out other possibilities, such as vortex pinning~\cite{kwon2018Magnetic} and the hysteresis effect~\cite{bothner2012Magnetic} (see SM for more details). Previous studies have shown that TLSs play an important role in determining the behaviors of superconducting resonators~\cite{kumar2008Temperature,gao2008Physics}. Given that this additional mechanism is related to the magnetic field, we propose a spin-flip TLS (sTLS) model to account for the anomaly.

\begin{figure}[tbp]
	\centering
	\includegraphics[width = 8.6 cm]{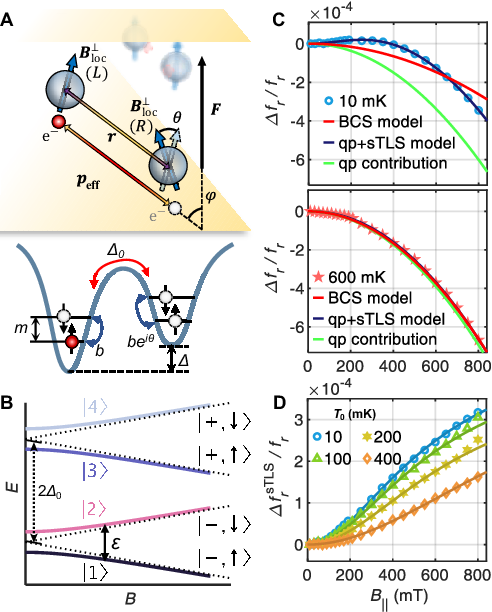}
	\caption[asd]{(A) Schematic illustration of the sTLS model with detailed explanations provided in the text. (B) Energy level diagram of the sTLS. (C) Measured $\Delta f_{r}/f_{r}$ at $T = 10$ and 600 mK, respectively, which are fitted by the BCS model (red curves), and the modified formula including the contributions of qp generation and sTLSs (blue curves). The green curves highlight the contributions of qp generation from the modified formula. (D) Extracted $\Delta f_{r}^{\rm sTLS}/f_{r}$ at different temperatures $T_0$, which are well-fitted by Eq.~(\ref{eq:5}). Here, the data are obtained by subtracting the fitting results of the BCS model at 600 mK.}
	\label{fig3}
\end{figure}

As shown in Fig.~\ref{fig3}A, two adjacent defects (labeled as big blue balls in the upper panel) separated by a distance of $r$ can form an in-situ double potential well to trap an electron (labeled as the small red ball) in either left or right well with the corresponding state labeled as $|L\rangle$ or $|R\rangle$. This system is characterized by the energy difference $\Delta$ between the two wells and the inter-well electron tunneling rate $\Delta_0$ (see the lower panel). Such a spin-independent system forms a single conventional TLS (cTLS), with its eigenstates denoted as orbital states $|\pm\rangle$~\cite{anderson1972Anomalous, phillips1972Tunneling}. When an external magnetic field $B$ is applied, the spin Zeeman effect splits each orbital state by an energy separation of $m=g\mu_{\rm B}B$, resulting in four eigenstates of $|\pm,\uparrow\!\!(\downarrow)\rangle$ (black dashed lines in Fig.~\ref{fig3}B). However, when the spin exchange interaction is absent, tunneling can occur only between states with conserved spins (for example, from $|-,\uparrow\rangle$ to $|+,\uparrow\rangle$)~\cite{polishchuk2005Effect}. Therefore, cTLS is not affected by the magnetic field, so that its response is unchanged when the magnetic field is varied.

In contrast, our proposed sTLS model considers the defects to be magnetic, thus inducing the spin exchange interaction~\cite{sereda2007Resonant,polishchuk2005Effect,desousa2007Danglingbond, pal2018YuShibaRusinov, shiba1968Classical}. Such an interaction between a magnetic defect and an electron can be classically treated as a local magnetic field $\bm{B}_{\rm loc}$ acting on the electron~\cite{shiba1968Classical}. Here, we only consider the component of $\bm{B}_{\rm loc}$ that is perpendicular to the external magnetic field (denoted as $\bm{B}_{\rm loc}^{\perp}$), as it induces spin flips of the electron~\cite{coey2021Handbook}. Notably, the spin orientations of the two adjacent magnetic defects are typically different. Therefore, the directions of $\bm{B}_{\rm loc}^{\perp}$ at each well also differ, with a relative angle denoted as $\theta$. This inhomogeneity of $\bm{B}_{\rm loc}^{\perp}$ results in an effective spin-orbit coupling, mixing the orbital states and the spin states~\cite{benito2017Inputoutput}. Without loss of generality, we define the direction of $\bm{B}_{\rm loc}^{\perp}$ in the left well as the x-axis and consider the case of $\Delta = 0$. Thus, the Hamiltonian of the system can be expressed as:
\begin{equation}\label{eq:1}
	H_{\rm sys}=\dfrac{1}{2}(m\sigma_{z}+2\Delta_{0}\tau_{x})+b(|L\rangle\langle L|\sigma_{x}+|R\rangle\langle R|\sigma_{\theta}).
\end{equation}

Here, $\sigma_{\alpha}\ (\tau_{\alpha})$ denotes the Pauli operator in the spin (position) space, and $\sigma_{\theta}=\cos\theta\sigma_{x}+\sin\theta\sigma_{y}$. $b$ is the spin-flip rate caused by $\bm{B}_{\rm loc}^{\perp}$. The eigenstates of $H_{\rm sys}$ are denoted as $|j\rangle$ for $j = 1$ to 4, with corresponding eigenenergies $E_j$ from low to high (see Fig.~\ref{fig3}B). Due to the small population of the two highest energy levels at low temperatures, only $|1\rangle$ and $|2\rangle$ are considered to participate in the formation of the sTLS. As shown in Fig.~\ref{fig3}B, the sTLS energy $\varepsilon$ is expressed as $E_2-E_1$. It is no longer dominated by $\Delta_{0}$ like cTLS but is related to $b$ and $\theta$, and increases with increasing $B$.

Next, we consider the influence of the sTLS on the resonator, which is mediated by the electric-dipole interaction between the sTLS and the microwave electric field $\bm{F}$~\cite{phillips1987Twolevel,gao2008Physics}. The initial electric dipole moment $\bm{p}$ originates from the inter-well transitions of the electron, given by $\bm{p}\!\sim\!{\rm e}\bm{r}$~\cite{agarwal2013Polaronic}. Due to the effective spin-orbit coupling induced by magnetic defects, the transition of sTLS ($|1\rangle\!\leftrightarrow\!|2\rangle$) also involves a change in the orbital component. Thus, its effective electric dipole moment operator $\bm{p_{\rm eff}}$ is determined by $\bm{p}$ and the spin-orbit mixing angle $\Gamma$, resulting in $\bm{p_{\rm eff}}=\bm{p}\sin\Gamma\sigma'_{x}$. Here, $\sin\Gamma \approx(b\sin\frac{\theta}{2})(\Delta_0\!-\!\frac{\varepsilon^2}{4\Delta_0})^{-1}$ (valid when $b\sin\frac{\theta}{2}\ll2\Delta_0-\varepsilon$), and $\sigma'_{x}$ is the Pauli x operator in the sTLS space. Consequently, the interaction Hamiltonian can be expressed as $H_{\text{int}} = p_{\rm eff}F\cos\varphi$, where $\varphi$ is the angle between $\bm{p_{\rm eff}}$ and $\bm{F}$, as illustrated in Fig.~\ref{fig3}A.

The energy change caused by $H_{\rm int}$ associated with $\bm{F}$ can be described by an additional electrical susceptibility~\cite{gao2008Physics}. Similar to the standard cTLS theory, we calculate the expression of the electrical susceptibility as
\begin{equation}\label{eq:2}
	\chi\!=\!\tanh{\!\dfrac{\varepsilon}{2k_{\rm B}T}\!}\left(\!\dfrac{2\varepsilon}{\varepsilon^2-(hf_r)^2}\!\right)\!\!({\rm e}r\!\cos\!\varphi)^2\!\!\left(\!\dfrac{b\sin\frac{\theta}{2}}{\Delta_0\!-\!\frac{\varepsilon^2}{4\Delta_0}}\!\right)^{\!\!\!2}\!.
\end{equation}

The contribution of all sTLSs to the electrical susceptibility will modify the dielectric function $\epsilon_{m}$ of the host material, thereby altering $f_r$. This modification in $\epsilon_{m}$ is expressed by integrating $\chi$ over relevant parameters:
\begin{equation}\label{eq:3}
	\epsilon_{\rm sTLS}(B,T)\!=\!\!\!\iiiint{\!\!\chi P_{\Delta_{0}} P_{\cos\varphi}P_bP_\theta}{\rm d}\Delta_{0}{\rm d}\!\cos\!\varphi{\rm d}b{\rm d}\theta,\!
\end{equation}where $P_{\langle\cdots\rangle}$ is the distribution of the corresponding parameter ($\langle\cdots\rangle=\Delta_{0}$, $\cos\varphi$, $b$ or $\theta$). 

As a parameter in the standard cTLS model, the distribution of $\Delta_0$ is widely accepted as $P_{\Delta_0}\sim n_{\rm sTLS}/\Delta_0$~\cite{phillips1987Twolevel}, where $n_{\rm sTLS}$ is the volume density of sTLSs. We assume that the direction of the electric dipole moment $\bm{p_{\rm eff}}$ is randomly distributed, which results in a uniform distribution of $P_{\cos\varphi}$. $P_b$ could be affected by the microscopic origin and structure of the magnetic defects~\cite{drera2012Electronic,hong2006Roomtemperature,zhou2009Origin}, the interaction mechanism~\cite{desousa2007Danglingbond,shiba1968Classical}, or the status of the magnetic domains~\cite{matsumoto2001RoomTemperature}; hence, we cannot accurately define it. However, we have verified through numerical calculations that it is feasible to treat $b$ as a constant $b_0$ to represent its average effect, that is, $P_b=\delta(b\!-\!b_0)$. The distribution of $\theta$ is related to the defect--defect interaction $J$, and can be expressed as $P_\theta\!=\![2k_{\rm B}T\!\sinh(\frac{J}{k_{\rm B}T})/J]^{-1}\!\sin(\theta) e^{\!\frac{-J\cos\theta}{k_{\rm B}T}}\!$~\cite{coey2021Handbook}. The sign of $J$ determines whether $\theta$ tends to be zero or $\pi$.

Note that different types of magnetic defects may have different $J$, leading to the formation of sTLSs with varying properties. In our samples, the substrate is treated with buffered oxide etching before sputtering, which has been shown to significantly remove magnetic defects at the substrate--metal and the substrate--vacuum interfaces~\cite{desousa2007Danglingbond,jayaraman2024Lossa}. Consequently, the magnetic defects, with their presence evidenced by the observed electron spin resonance, are likely to concentrate at the metal--vacuum interface of TiN. This interface is known to have extensive magnetic defects originating from titanium oxide and oxynitride~\cite{driessen2012Strongly, saveskul2019Superconductivity}. In these materials, room-temperature ferromagnetism is widely observed, corresponding to a negative $J$ of $\sim\!\rm meV$ well above our experimental temperatures~\cite{drera2012Electronic,hong2006Roomtemperature,zhou2009Origin}. Therefore, $\theta$ should be close to zero, which makes $\varepsilon\approx\sqrt{m^2\!+\!4b^2}$.

Substituting these distributions into Eq.~(\ref{eq:3}), we can obtain the final expression of $\epsilon_{\rm sTLS}$:

\begin{equation} \label{eq:4}
	\epsilon_{\rm sTLS}(B,T)\!=\!\dfrac{At^2\sqrt{(vB)^2\!+\!t^2}}{(vB)^2\!+\!t^2\!-\!(\!\frac{hf_r}{k_{\rm B}}\!)^2}\tanh{\dfrac{\sqrt{(vB)^2+t^2}}{2T}},\!\!\!
\end{equation}where $t\!=\!2b_0/k_{\rm B}$ represents the average zero-field energy of sTLSs, and $v\!=\!g\mu_{\rm B}/k_{\rm B}$ converts $B$ into the temperature dimension. $A$ is the integration coefficient which is positively related to $n_{\rm sTLS}$. Thus, the sTLS-induced fractional frequency shift can be calculated as~\cite{gao2008Physics}
\begin{equation} \label{eq:5}
	\dfrac{\Delta f_{r}^{\rm sTLS}(B,T)}{f_{r}}\!\approx\!-G\dfrac{\epsilon_{\rm sTLS}(B,T)\!-\!\epsilon_{\rm sTLS}(0,\!T)}{2\epsilon_{m}}\!\triangleq\!h(B,T),
\end{equation}where $G$ is the filling factor which represents the proportion of the electric field in the sTLS host material. Using COMSOL simulation, we estimate $G\sim10^{-4}$.

Therefore, combined with the standard BCS theory of qp generation, the total fractional frequency shift is expressed as $\Delta f_{r}(B,T)/f_{r}=k B^{2}+h(B,T)$. We use this formula to fit the experimental results with three fitting parameters, $k$, $W=GA/(2\epsilon_{m})$, and $t$. As shown in the upper panel of Fig.~\ref{fig3}C, the data measured at 10 mK, which cannot be well-fitted using the pure BCS model (red curve), are quantitatively reproduced by our modified formula (blue curve). We can separate the qp contribution (green curve) from this fitting, with $k = (-0.94\pm0.03)\times10^{-9}\ \rm mT^{-2}$.

As described by Eq.~(\ref{eq:2}), $\chi$ decreases with increasing $T$, similar to the thermal saturation of cTLS~\cite{anderson1972Anomalous, phillips1972Tunneling}. Therefore, it is expected that at sufficiently high $T$, the sTLSs contribution will be suppressed, causing $\Delta f_{r}/f_{r}$ to be dominated by the commonly observed mechanism of qp generation. As shown in the lower panel of Fig.~\ref{fig3}C, we investigate $\Delta f_{r}/f_{r}$ at $T=600$ mK. At such a high temperature, the anomaly in the frequency--field relationship almost disappears. Fitting the experimental results using the pure BCS model (red curve) or the modified formula (blue curve) shows negligible differences, suggesting the thermal saturation of sTLSs. We extract $k'=(-1.03\pm0.01)\times10^{-9}\ \rm mT^{-2}$ from the red curve. Previous studies have shown that field-induced qp generation is independent of $T$~\cite{healey2008Magnetic,tinkham1996Introduction,foshat2023Characterizing}. Indeed, we find that $k'$ is close to $k$ extracted from the 10 mK data. This indicates the accuracy of the modified formula in describing an additional mechanism (sTLS) superimposed on the traditional mechanism of qp generation.

We have taken a closer look at the sTLSs contributions by subtracting the qp contributions (fitting results $k'B^2$ from the BCS model at 600 mK) in the total frequency shifts at various $T_0$. As shown in Fig.~\ref{fig3}D, the sTLSs contributions from 10 to 400 mK show a similar increasing trend with the magnetic field and become less pronounced with increasing $T$. All these behaviors are well-fitted by Eq.~(\ref{eq:5}) with fitting parameters $W$ and $t$, as indicated by the corresponding solid curves. To demonstrate reproducibility, we repeat such experiments in four resonators with different sample parameters and field directions, all of which show similar behaviors and can be well-fitted (see SM).

\begin{figure}[tbp]
	\centering
	\includegraphics[width =1\linewidth]{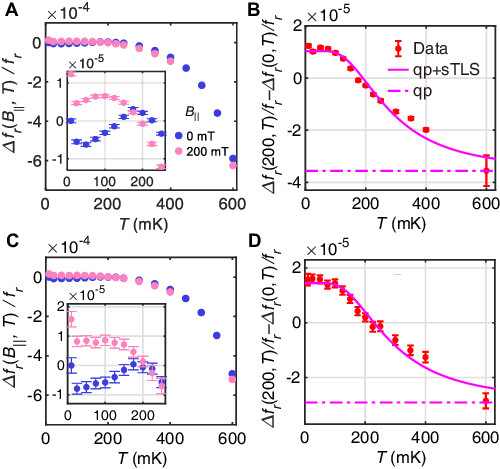}
	\caption[asd]{Measured $\Delta f_{r}/f_{r}$ of resonators (A) A1 and (C) A4 as a function of $T$ at $B_\parallel$ of 0 mT and 200 mT, respectively, with the low-temperature regime zoomed in the insets. (B) and (D) show differences between the curves of 200 mT and 0 mT in (A) and (C), respectively, which are well-fitted by $kB_0^{2}+h(B_0, T)$. The dash-dotted lines correspond to the qp contributions of $kB_0^{2}$.}
	\label{fig4}
\end{figure}

Moreover, to better demonstrate the temperature-dependent behaviors of sTLSs, we measure the evolution of $\Delta f_{r}/f_{r}$ of A1 and A4 as a function of $T$, under $B_\parallel$ of 0 mT and 200 mT. As shown in Figs.~\ref{fig4}A and \ref{fig4}C, both curves show an overall decreasing frequency trend at higher temperatures due to temperature-induced qp generation~\cite{kumar2008Temperature}. In the low-temperature regime (inset of Figs.~\ref{fig4}A and \ref{fig4}C), where the contribution of TLSs (including sTLSs discussed here and cTLSs originating from nonmagnetic disorders) is pronounced, the signature of a detailed pattern is visible. To better focus on the sTLSs contribution, we calculate the difference between the two curves. This not only subtracts the background which is independent of $B$ (for example, the influence of temperature-induced qp generation and cTLSs), but also suppresses the influence of the frequency drift accumulated during long-term measurements (note that for both resonators, data points at the base temperature exhibit jumps with respect to other data points, which are collected continuously a few days later, while these jumps are absent after taking the difference). Figures~\ref{fig4}B and \ref{fig4}D show the extracted difference. Interestingly, the experimental results exhibit an obvious temperature dependence, which deviates from the constant value of $kB_0^2$ (where $B_0=200$ mT, shown as the dash-dotted lines) predicted by the BCS model. However, the data can be well-fitted by our sTLS model, especially in the low-temperature regime. This result indicates the pronounced influence of sTLSs and its evolution upon thermal saturation.

We first discuss the obtained fitting parameters of our sTLS model. The fitting parameters $W$ and $t$ from all field-dependent and temperature-dependent data are almost of the same order of magnitude (as summarized in SI). $t$ mostly falls within the range of 300 – 600 mK. This means that the average zero-field energy of sTLSs is in the range of 6.3 – 12.5 GHz in our system, affecting $f_r$ mainly through nonresonant interactions. These sTLSs are formed by tunneling of electrons according to our model. This is intriguing since tunneling electrons are usually believed to contribute at a much higher frequency range, due to their small mass, thus a large tunneling rate $\Delta_0$ (typically 1 meV – 1 eV depending on the tunneling distance)~\cite{agarwal2013Polaronic}. However, when spin-orbit coupling is present, it mixes charge tunnelings and spin flips, causing the sTLS energy to be dominated by spin-flip rate $b_0$ (estimated to be $\sim10\ \mu$eV in our experiment) instead of $\Delta_0$. This understanding differs from previously proposed mechanisms, such as dressing via phonon states~\cite{agarwal2013Polaronic} or virtual tunneling processes~\cite{chakravarty1987Effect}.

From the average fitting value of $W$, we estimate the magnetic defect density $n_d$ to be approximately $10^{17}\ \rm m^{-2}$ (see SI). This value is consistent with those observed in realistic quantum devices~\cite{saveskul2019Superconductivity,koch2007Model,sendelbach2008Magnetism,degraaf2018Suppression,jayaraman2024Lossa}. Note that such magnetic defects have been shown to widely exist at various metal or substrate surfaces, such as TiN~\cite{saveskul2019Superconductivity}, Al~\cite{kumar2016Origin}, NbN~\cite{degraaf2018Suppression,jayaraman2024Lossa}, Nb~\cite{proslier2008Tunneling}, and other materials~\cite{vranken1988Enhanced,degraaf2017Direct}. Therefore, our model can be extended to those material systems in which different sTLS spectra are expected with different defects' magnetic properties.

More importantly, our model can be used to guide engineering the noise and loss of quantum devices through controlling magnetic fields. For example, considering systems with paramagnetic defects, an external magnetic field can easily polarize defect spin orientations, decreasing the inhomogeneity of the local magnetic field and thus the generated effective spin-orbit coupling. This directly reduces $p_{\text{eff}}$ of the sTLSs, decoupling them from microwave electric fields and other TLSs. This understanding may explain recent experiments where increasing magnetic field suppresses charge noise and loss in quantum devices~\cite{degraaf2017Direct,khalifa2023Nonlinearity,zhang2024AcceptorInduced,barone2018Kondolike}. Moreover, it also explains a more pronounced loss suppression with magnetic fields observed under an intermediate microwave drive power~\cite{borisov2020Superconducting}. This is probably because this intermediate power saturates cTLSs while leaving sTLSs unsaturated due to their smaller $p_{\text{eff}}$ (see SI). At the meantime, higher magnetic fields may degrade quantum coherence through, for example, mobility reduction~\cite{muoi2024Effect} or Cooper pair breaking~\cite{samkharadze2016HighKineticInductance}. Therefore, the competition between these mechanisms should result in an optimal magnetic field that minimizes decoherence. This optimal field can be further engineered to align with the sweet spot of spin operations~\cite{benito2019Optimized} to improve the performance of spin qubits~\cite{connors2019Lowfrequency, yoneda2018quantumdot}.

In conclusion, we demonstrate anomalous frequency–field and frequency–temperature relationships in TiN ultra-thin film resonators. These observations are explained by our sTLS model involving spin flips, which originate from magnetic defects at interfaces. Our results reveal the spin-dependent behaviors of TLSs, which are in the frequency range commonly used in quantum devices (1–10 GHz), thus having great implications for the improvement of their decoherence.

\begin{acknowledgments}
	This work was supported by the National Natural Science Foundation of China (Grants No. 92265113, No. 12074368, No. 12034018, and No. 12274397), by the Natural Science Foundation of Jiangsu Province (Grant No. BK20240123), by the Innovation Program for Quantum Science and Technology (Grant No. 2021ZD0302300). This work was partially carried out at the USTC Center for Micro- and Nanoscale Research and Fabrication.
\end{acknowledgments}

\providecommand{\noopsort}[1]{}

\end{document}